\documentclass[useAMS,usenatbib,onecolumn]{mn2e}
\usepackage{amsmath}
\usepackage{epsfig,color}
\usepackage{dcolumn}
\usepackage[utf8]{inputenc}
\usepackage{graphics,graphicx}
\usepackage{epstopdf}
\usepackage{float}
\usepackage{subfigure}
\usepackage[para]{threeparttable}
\usepackage{url}
\usepackage{color}

\newcommand{\3}{$_{3}$}

\newcommand{\lnr}{^{\ell}}

\newcommand{\cm}{cm$^{-1}$}
\newcommand{\icm}{cm\textsuperscript{-1}}

\newcommand{\ai}{\textit{ab initio}}

\newcommand{\micro}{$\mu$m}

\newcommand{\p}{^\prime}
\newcommand{\pp}{^{\prime\prime}}

\title[ExoMol line lists VIII: H$_2$CO]{ExoMol line lists VIII: A variationally computed line list for hot formaldehyde}

\date{\today}

\author[Al-Refaie et al]{\large Ahmed F. Al-Refaie,  Andrey Yachmenev, Jonathan Tennyson and Sergei N. Yurchenko,\\
Department of Physics and Astronomy, University College London, Gower Street, WC1E 6BT London, UK}

\date{Accepted XXXX. Received XXXX; in original form XXXX}

\pagerange{\pageref{firstpage}--\pageref{lastpage}} \pubyear{2013}

\begin{document}

\label{firstpage}

\maketitle

%Work on splittingup the long sentences
\begin{abstract}
  A computed line list for formaldehyde, H$_2{}^{12}$C$^{16}$O, applicable to
  temperatures up to $T=1500$~K is presented.  An empirical potential energy
  and {\it  ab initio} dipole moment surfaces are used as the input to
  nuclear motion program TROVE. The resulting line list, referred to as
  \textit{AYTY}, contains 10.3 million rotational-vibrational states and
  around 10 billion transition frequencies. Each transition includes
  associated Einstein-$A$ coefficients and absolute transition intensities,
  for wavenumbers below 10~000 cm\(^{-1}\) and rotational excitations up to
  \(J=70\). Room-temperature spectra are compared with laboratory
  measurements and data currently available in the HITRAN database. These
  spectra show excellent agreement with experimental spectra and highlight
  the gaps and limitations of the HITRAN data.  The full line list is
  available from the CDS database as well as at \url{www.exomol.com}.
\end{abstract}

% Create a new 1st level heading
\section{Introduction}

Formaldehyde, H$_2$CO, is  a poisonous molecule in the aldehyde group.
On Earth it plays a part in troposphere chemistry dynamics as the main source of OH
via photo-dissociation and is formed from photo-oxidation in the atmosphere or
through the incomplete burning of biomass \citep{00WaXXXX.H2CO}.
Traces of formaldehyde have tentatively been detected in the martian atmosphere \citep{93KoAcKr.H2CO}
where it is believed to be derived from the oxidation of methene (C$_{2}$H$_{4}$) \citep{13ViMuNo.H2CO}.

Formaldehyde was the first polyatomic molecule to be detected in the interseller medium (ISM)
\citep{70ZuBuPa.H2CO} and is extremely abundant \citep{76LaXXXX.H2CO}.
This has made it useful in investigating the isotope composition of carbon in the
galaxy \citep{74ZuBuPa.H2CO}. The proposed mechanism of production is via the
successive hydrogenation of CO \citep{02DaXXXX.H2CO} on icy grain mantles:
\begin{eqnarray}
\nonumber
  {\rm H} + {\rm CO} &\rightarrow& {\rm HCO }\\
  {\rm H} + {\rm HCO}  &\rightarrow& {\rm H}_{2}{\rm CO}.
\end{eqnarray}
Further hydrogenation produces methanol through an intermediate methyl radical
H+H$_{2}$CO$\rightarrow$ CH$_{3}$O$\rightarrow$ H+ CH$_{3}$O $\rightarrow$ CH$_{3}$OH.
Common reactions include that with ammonia which produces amines \citep{02ScXXXX.H2CO} and
polymerisation with other H\(_{2}\)CO  molecules. As a result, formaldehyde is believed to be
the major precursor for the formation of complex organic molecules in the ISM that include
interstellar glycolaldehyde \citep{00HoLoJe.H2CO} and amino acids \citep{02ScXXXX.H2CO}.

Formaldehyde's astrophysical relevance does not end in the ISM. Recently,
it has been detected in comets \citep{92BoCrxx.H2CO}, such as 103P/Hartley 2 \citep{11DeVeLi.H2CO},
C/2007 N3\citep{11ViMuDi.H2CO} and Hale-Bopp \citep{06StAnMa.H2CO}, where it is thought to originate
 from the degredation of polyoxymethylene \citep{01CoGaBe.H2CO}.
It is also present in protoplanetary discs around low mass young stars (Taurus-Auruga Class I/II)
 \citep{10ObQiFo.H2CO,09ZaKeWa.H2CO,14SaFoWa.H2CO} as circumstellar ice with an
abundance ratio of \(\approx 2\%\) compared to the more ubiquitous water-ice.

%-----------------------------------
Because of H\(_{2}\)CO's role as a precursor to complex organic molecules,
it is considered a possible biomarker. The RNA world hypothesis suggests an early
Earth with a CO$_{2}$, H$_{2}$O and N$_{2}$ rich atmosphere \citep{13NeKiBe.H2CO}.
Illuminating this mix with ultraviolet (UV)
radiation should lead to a large amount of formaldehyde being fixed in the atmosphere
before being deposited into the prebiotic oceans \citep{13NeKiBe.H2CO}.
Alternatively, the source of prebiotic chemical compounds may be derived without need
of illuminating UV radiation via glancing icy body impacts \citep{13GoTaxx.H2CO}.
Such impacts would produce shock-compression conditions that lead to formation of HCN molecules.
These HCN molecules can be hydrolyzed to form formaldehyde and from there produce amino acids.
Thus a planet rich in formaldehyde may indicate one undergoing the stages of pre-life.
%----------------------------------

Finally, formaldehyde masers  \citep{80FoGoWi.H2CO,92PrSnBa.H2CO} are a reliable and proven tracer for high-density environments
such as star-forming regions in galaxies due to its ubiquity and large number of long
wavelength transitions \citep{08MaDaMe.H2CO}. Currently, there are 19 extragalactic
sources \citep{08MaDaMe.H2CO} of these masers including IRAS 18566 + 0408, which is
notable for detection of the first H\(_{2}\)CO maser flare  \citep{07ArHoSe.H2CO}.
Formaldehyde masers  (and maser flares) have mostly been observed via the
\(1_{10}\rightarrow1_{11}\) and \(2_{11}-2_{12}\) $K$-doublet transitions at 6.1 cm and
2.2 cm respectively.
%its cm, its 4.83 GHz which corresponds to about 0.16cm-1 or 6.1 cm

The wide-range of interactions in atmospheric, terrestrial, astrophysical and
astrobiological phenomena makes formaldehyde a relevant molecule in the chemistry of
exoplanets and their atmospheres. Therefore a complete, high-resolution,  line list for
H\(_{2}\)CO should provide an important aid for characterisation and modelling of
formaldehyde. These considerations led us to study formaldehyde as part of  the ExoMol
project \citep{jt528},  which aims to produce comprehensive molecular line lists for
studies of the atmospheres of exoplanets and cool stars.

%\red{AHMED, DO ME MENTION HERE THE PAPER WITH INTENSITIES IN THE HIGHER WAVELENGTH REGION?}
High-resolution, room-temperature formaldehyde spectra have been well-studied
in the laboratory
\citep{75JoMcxx.H2CO,79BrHuPi.H2CO,82NaTaKo.H2CO,87NaDaRe.H2CO,88ClVaxx.H2CO,89ReNaDa.H2CO,96PoBrCa.H2CO,03ThCaRi.H2CO};
the early work was reviewed by \citet{83ClRaxx.H2CO}.  Currently, the major
source of publicly available spectroscopic data on H\(_{2}\)CO is the HITRAN
database \citep{jt557} which has recently been updated to include long-wavelength data
from the CDMS database \citep{05MuScSt.db}. The spectral regions covered in the
database are 0 -- 100 cm\(^{-1}\), 1600 -- 1800 cm\(^{-1}\)
\citep{09PeJaTc.H2CO} and the 2500 -- 3100 cm\(^{-1}\) \citep{09PeJaTc.H2CO} at
up to 10\(^{-29}\) cm/molecule sensitivity for $T$=296 K. However, this
accounts for only 40~000 transitions extending up to \( J = 64 \) and covers
only four of the six fundamental vibrational bands as well as the ground state
rotational spectrum. This deficiency arises from an apparent lack of absolute
intensities in the 100 -- 1600 cm\(^{-1}\) range. Additional observed
transitions are available \citep{09PeJaTc.H2CO} and include line positions
\citep{03PeKeFl.H2CO,07TcPeLa.H2CO,07ZhGaDe.H2CO}, and intensities
\citep{03PeKeFl.H2CO,06PeVaDa.H2CO,06FlLaSa.H2CO} of some of the fundamental
bands and hot bands \citep{94ItNaTa.H2CO,06PeBrUt.H2CO,09MaPeJa.H2CO}.
%(\(\nu_{4}-\nu_{4}\), \(\nu_{1}+\nu_{4}-\nu_{4}\) and \(\nu_{5}+\nu_{4}-\nu_{4}\)) \citep{06PeBrUt.H2CO} with band centers in the 2760-2860 cm\(^{-1}\) region.
The incompleteness and low rotational excitations available in HITRAN limits
the applicability of this data for temperatures above 300 K. The theoretical
spectra presented in this paper aims to provide a more complete and accurate
picture of the spectra of formaldehyde up to 10~000 cm\(^{-1}\) and for
temperatures up to 1500 K. Our line list should therefore be useful for
modelling higher temperature environments as well as studies on non-LTE
transitions such as those observed in masers.

Theoretically, electric dipole transition intensities of H\2CO were studied by
\citet{96LuCoFr.H2CO} and \citet{09CaStAm.H2CO}; see also the review by
\citet{13Yuxxxx.method}. \citet{96LuCoFr.H2CO} used an \textit{ab initio}
MP2/6-311G** DMS to simulate the photoacoustic spectrum of high C-H stretching
overtones of H$_{2}$CO.  \citet{09CaStAm.H2CO} generated an \ai\ couple-cluster
CCSD(T)/aug-cc-pVTZ dipole moment surface (DMS) for H\2CO; they used an
effective charges representation to compute (relative) rovibrational line
intensities for H$_{2}$CO reproducing the HITRAN data \citep{jt453} with
reasonable agreement. \citet{96PoBrCa.H2CO} computed an \ai\ DMS using the
QCISD/6-31111G(d,p) level of theory and presented it as an expansion. %using a
%bisector and Eckart frame.

Despite these works there is no comprehensive line list for
formaldehyde available in the literature. The goal of this work is to bridge this gap. We
use the variational program TROVE \citep{07YuThJe.method} in conjunction with an initial
potential energy surface (PES) obtained
`spectroscopically' by \citet{11YaYuJe.H2CO} and a new \ai\ dipole moment surface (DMS) for formaldehyde and
generate an extensive line list for H$_2{}^{12}$C$^{16}$O applicable for the
temperatures up to $T=1500$~K. In the following H$_2$CO and formaldehyde will
refer to the main isotopologue H$_2{}^{12}$C$^{16}$O.

\section{Method}
\label{s:method}

\subsection{Background}

H$_{2}$CO is a prolate asymmetric top molecule that belongs to the
C\(_{2v}\) molecular symmetry group \citep{98BuJexx.method}. The group
has four irreducible representations \(A_{1}\), \(A_{2}\), \(B_{1}\)
and \(B_{2}\). Once the H atom nuclear spin is taken into account the
`para' $A$ representations are singly degenerate and the 'ortho' $B$
representations are triply degenerate.  As H$_{2}$CO has four atoms,
it has six vibrational modes; Table \ref{tab:vibcoord} shows the
vibrational modes and their corresponding symmetries, band centers and
descriptions.  Coriolis interactions occur strongly between the
\(\nu_{4}\) and \(\nu_{6}\) modes, and weakly between the \(\nu_{3}\)
and \(\nu_{4}\) modes \citep{71NaYoXX.H2CO} which couples their energy
levels and wavefunctions. This manifests itself in the \(\nu_{3}\),
\(\nu_{4}\) and \(\nu_{6}\) mode interaction as overlapping bands
which make these three bands difficult to distinguish from each other.
%\red{CHECK IF THERE ANY THEORETICAL LINE LISTS EXIST IN THE LITERATURE AND CITE IF THEY DO OTHERWISE: }

\begin{table}
  \centering
  \caption{Vibrational modes and observed band centres in cm$^{-1}$ by \citet{95CaPiHa.H2CO}.}
  \label{tab:vibcoord}
    \begin{threeparttable}
     \begin{tabular}{c*{3}l}
\hline\hline
        Mode              & Band Centers %\tnote{1,2}
& Symmetry & Description \\
        \hline
         \(\nu_{1}\)  & 2782.46 & \(A_{1}\) & symmetric C-H stretching   \\
         \(\nu_{2}\)  & 1746.01 & \(A_{1}\) & C-O stretching  \\
         \(\nu_{3}\)  & 1500.18 & \(A_{1}\) & symmetric O-C-H bending \\
         \(\nu_{4}\)  & 1167.26 & \(B_{1}\) &  out-of-plane bending \\
         \(\nu_{5}\)  & 2843.33 & \(B_{2}\) &  asymmetric C-H stretching\\
         \(\nu_{6}\)  & 1249.10 & \(B_{2}\) & asymmetric O-C-H bending\\
\hline
         \end{tabular}
    \end{threeparttable}
%    \begin{tablenotes}
%      \item [1] \cm
%      \item [2] \citet{95CaPiHa.H2CO}
%    \end{tablenotes}
\end{table}

\subsection{Potential energy surface}

Full details of their PES calculation are given by \citet{11YaYuJe.H2CO},
so only a brief summary is presented here. The initial PES was computed
\textit{ab initio} using the CCSD(T)/aug-cc-pVQZ theory. %\citep{85UrNoCo.ai}
%as implemented in the MOLPRO package \citep{12WeKnKn.methods}.
Variational calculations with this surface give a root-mean-square
(rms) error of 5.1 cm\(^{-1}\) for the fundamental band centers.
\cite{11YaYuJe.H2CO} refined the \ai\ PES using a \( V^{\prime} = V +
\Delta V\) formulation where \(V\), \(\Delta V\) and \(V^{\prime}\)
are the original \textit{ab initio}, \textit{correction} and
\textit{refined} PES respectively.
%The eigenfunctions of the original
%`unperturbed' Hamiltonian, \(H\), are used as the basis functions for
%the extended Hamiltonian \(H' = H + \Delta V\) where \( \Delta V\) is
%the only off-diagonal part of the Hamiltonian. \(\Delta V\) is
%expanded and the correction terms are obtained via a least-squares fit
%using TROVE.
The eigenfunctions of the original \ai\ Hamiltonian, \(H = T + V\), are used as
the basis functions for the extended Hamiltonian \(H' = H + \Delta V\) where \(
\Delta V\) is typically small and almost diagonal correction. \(\Delta V\) is
expanded in Taylor series and expansion coefficients are obtained in a
variational least-squares fit to a high-resolution spectroscopic data using
TROVE. The rms error against experimental energy levels with \(J \leq 5\) of
this semi-empirical PES, called H2CO-2011, is 0.04 cm\(^{-1}\).

\begin{figure} \begin{center}
    \centering
    \epsfxsize=15.0cm \epsfbox{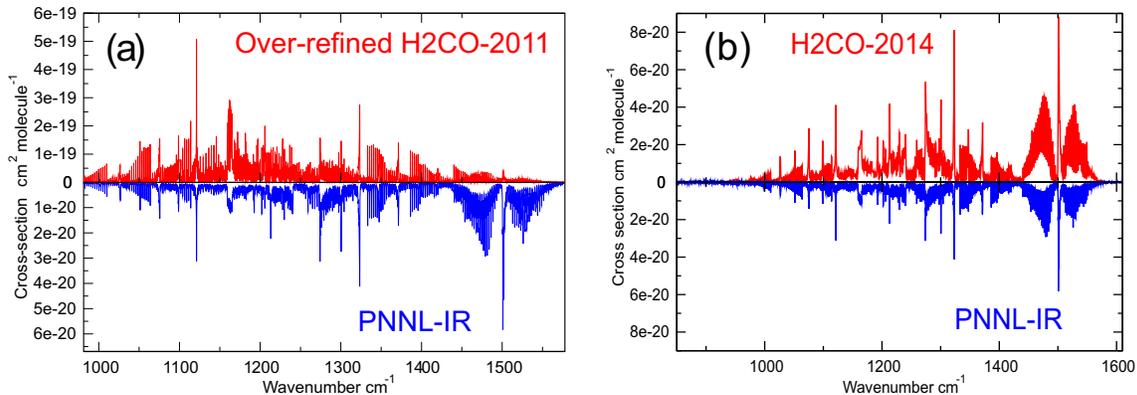}
    \caption{Overview of the \(\nu_3\), \(\nu_4\) and \(\nu_6\) region between different PES used in production and PNNL-IR~\protect\citep{PNNL} cross-sections:
     (a) represents the over-refined PES (b) the 'better' refinement.
     Note the difference in $y$-axis scaling in (a) to highlight structural features.}
    \label{fig:comp_pes}
    \end{center}
\end{figure}

It should be noted however that the excellent accuracy of the refined PES
caused serious problems with the absolute intensities of some bands. The
intensities based on the initial, less accurate \ai\ PES, the intensities of
the \(\nu_3\), \(\nu_4\) and \(\nu_6\) bands were observed to agree much better
with the experiment. Figure \ref{fig:comp_pes}(a) highlights this effect, where
an order of magnitude difference in absolute intensity and cross-section was
observed. Initially blamed on the DMS, it was soon discovered that the original
\ai\ PES did not have this problem.

In order to address this issue we have repeated the refinement process making
it less aggressive with careful observation of the transition moments. In the
present work we have also increased the size of the basis set (see discussion
below). We used the same fitting set of experimental energies as well as the
same functional form for PES. The rms deviation of the experimental term values
used in the PES fit ($J\le 5$) against the computed energies in AYTY is 0.18
cm$^{-1}$ (0.006~\cm\ for pure rotational term values). The potential
parameters as well as the associated Fortran 95 program are given as
supplementary material. The resulting line intensities have returned to quality
of the \textit{ab initio} levels as can seen in Figure \ref{fig:comp_pes}(b).

% making it an excellent starting point for the production of high-accuracy spectra.

\subsection{Variational computation}

The TROVE program suite \citep{07YuThJe.method} is employed to compute our
formaldehyde line list as well to perform the least-squares fit of the
\textit{ab initio} PES discussed above.  TROVE is designed to compute
variational ro-vibrational energy levels, associated eigenfunctions and
transition intensities for molecules of arbitrary structure. Variational
methods are often limited in their efficiency due to the need to diagonalize
increasingly large Hamiltonian matrices as the complexity of the molecule
increases. However with the improved computational power and parallelism of
modern CPUs, it is now feasible to solve ro-vibrational Schr\"{o}dinger
equations for polyatomic molecules. TROVE has been successfully used to produce
high-accuracy line lists and spectra for tetratomic molecules such as HSOH
\citep{10YaYuJe.HSOH}, NH\(_{3}\) \citep{jt500}, SO\(_{3}\) \citep{jt554},
PH\(_{3}\) \citep{jt556,jt592}, and SbH\3~\citep{10YuCaYa.SbH3}, as well as
recently the pentatomic molecule CH$_4$~\citep{jt564}.

%TROVE constructs and Taylor expands in generalised coordinates an approximate
%kinetic energy (AKE) operator using a recursive numerical scheme.
%Previous calculations for H\(_{2}\)O\(_{2}\) by \citet{jt553} show that computations
%using AKE for  non-linear molecules, such as H\(_{2}\)CO, converge as well as
%exact kinetic operator-based programs such as WAVR4 \citep{jt339} and,
%particularly for excited rotational states, at cost of significantly less
%computational time.
TROVE approximates the kinetic energy operator (KEO) by a truncated Taylor series expansion in generalized coordinates.
Previous calculations for H\(_{2}\)O\(_{2}\) by \citet{jt553} show that computations
using series representation of the KEO for non-linear molecules, such as H\(_{2}\)CO, converge as well as
exact KEO-based programs such as WAVR4 \citep{jt339} and,
particularly for excited rotational states, at cost of significantly less
computational time.
In this work we use a kinetic expansion order of \(6\) for
optimal convergence while providing reasonable computation times and similarly,
we use a potential energy expansion order of \(8\).

In TROVE, the primitive vibrational basis set is represented by a symmetrized product of
six one-dimensional vibrational functions $\phi_{n_1}(r_1\lnr)$, $\phi_{n_2}(r_2\lnr)$,
$\phi_{n_3}(r_3\lnr)$, $\phi_{n_{4}}(\theta_{1}\lnr)$, $\phi_{n_{5}}(\theta_{2}\lnr)$,
and $\phi_{n_{6}}(\tau)$, where \(n_{i}\) denotes the associated local mode vibrational
quanta, $\{r_1\lnr, r_2\lnr, r_3\lnr, \theta_{1}\lnr, \theta_{2}\lnr\}$ are linearized
versions~\citep{07YuThJe.method,98BuJexx.method} of the coordinates $\{r_{\rm CO},
r_{{\rm CH}_1}, r_{{\rm CH}_2}, \theta_{{\rm OCH}_1}$, and $\theta_{{\rm OCH}_2}\}$,
respectively, and $\tau$ is the dihedral angle between the OCH$_1$ and OCH$_2$ planes.
The functions $\phi_{n_i}(q_i)$ are obtained by solving the corresponding 1D
Schr\"{o}dinger equation~\citep{07YuThJe.method} for the vibrational motion associated
with the corresponding coordinate $q_i$ $\in$ $\{ r_1\lnr, r_2\lnr, r_3\lnr,
\theta_{1}\lnr, \theta_{2}\lnr, \tau \}$, with the other coordinates held fixed at their
equilibrium values, where the Numerov-Cooley
method~\citep{24Nuxxxx.method,61Coxxxx.method} is used. The direct product of the 1D
basis functions is contracted using the polyad condition:
\begin{equation}
%P = 2 (n_{1} + n_{4} ) + n_{2} + n_{3} + n_{5} + n_{6} \le P_{\rm max}.
P =  2 (n_{2} + n_{3}) + n_{1} +  n_{4} + n_{5} + n_{6} \le P_{\rm max},
\label{eq:polyad}
\end{equation}
which in terms of the normal mode quantum numbers $v_{i}$ reads
\begin{equation}
P = 2 (v_{1} + v_{5} ) + v_{2} + v_{3} + v_{4} + v_{6} \le P_{\rm max}.
\label{eq:polyad:normal}
\end{equation}
This polyad rule is based on the approximate relationship between the H\2CO fundamental frequencies (see Table~\ref{tab:vibcoord}):
%\red{CHECK}
\begin{equation}
\nu_{1} \approx \nu_{5}  \approx  2 \nu_2  \approx  2 \nu_3 \approx  2 \nu_4 \approx  2 \nu_6.
\label{eq:nu}
\end{equation}

The vibrational basis set is further optimized by solving four reduced
eigen-problems variationally for $\{q_1\}$, $\{q_2,q_3\}$, $\{q_4,q_5\}$, and
$\{q_6\}$ to produce four sets of wavefunctions $\Phi_{n_1}^{(1)}(q_1)$,
$\Phi_{n_2,n_3}^{(2,3)}(q_2,q_3)$, $\Phi_{n_4,n_5}^{(4,5)}(q_4,q_5)$, and
$\Phi_{n_6}^{(6)}(q_6)$, respectively. At the step~2 the pure vibrational
($J=0$) problem is solved variationally using the basis set constructed as a
symmetrized direct product of $\Phi_{n_i}^{(i)}$ ($i=1,6$) and
$\Phi_{n_j,n_k}^{(j,k)}$ ($j,k=2,3$ or $4,5$ ) contracted through the polyad
number condition (\ref{eq:polyad}) and symmetrized according to the C$_{2v}$(M)
molecular symmetry group.  In this work the basis set is truncated at \(P_{\rm
max} = 16 \) as the relative simplicity of the molecule means that this gives
well-converged results. The maximum polyad number \(P_{\rm
  max}\) restricts the number of combinations of
\(\phi_{n_{i}}^{(i)}\) and $\Phi_{n_i,n_j}^{(i,j)}$ for which \(P \leq P_{\rm
max} \).  The resulting eigenfunctions $\Psi_{i}^{J=0,\Gamma}$ obtained for
each C$_{2v}$(M) symmetry $\Gamma = A_1, A_2, B_1$ and $B_2$ together with the
symmetrized rigid rotor wavefunctions $|J,K,\tau_{\rm rot}\rangle$ form our
$J=0$ basis set representation~\citep{jt466}, where the ro-vibrational basis
functions are given as a direct product of $\Psi_{i}^{J=0,\Gamma}$ and
$|J,K,\tau_{\rm rot}\rangle$. Here $\tau_{\rm rot}$ is the rotational parity
defined by~\citet{05YuCaJe.NH3}, and $K$ is the projection of the angular
momentum on the body-fixed axis $z$. The latter is defined according with the
Eckart conditions~\citep{35Ecxxxx.method} and is oriented approximately along
the CO bond.
In C$_{2v}$(M) symmetry, $K$ and $\tau_{\rm rot}$ correlate with the customary $K_a$ and $K_c$ rotational quantum numbers as
\begin{equation}
K = K_{a},\quad \tau_{\rm rot} = \text{mod} (|K_a-K_c|,2).
\end{equation}
The vibrational part of the $J=0$ basis set is truncated using the
energy threshold of $hc \, 18\,000$ \icm and thus consists of 2310, 1531, 1688,
and 2112 functions for the $A_1, A_2, B_1$ and $B_2$ symmetries, respectively.

The resulting ro-vibrational Hamiltonian matrix in the $J=0$ representation
exhibits a block diagonal structure where each of four blocks represents an
irreducible representation $A_1, A_2, B_1$ or $B_2$ and can be diagonalised
independently. Each of these blocks
displays a band-diagonal structure %illustrated in Figure\ref{fig:blockmatrix}
whose bandwidth and length is determined by the
$J=0$ basis set size and the level of rotational excitation
respectively.
%This in-turn reduces
%the bandwidth and block-size of the matrices in order to make better
%use of computational resources.
%In this work we truncate at
%\(P_{max} = 16 \)  as the relative simplicity of the
%molecule means that this gives well-converged results.

In generating our line list we employed an upper eigenvalue limit of 18~000
cm\(^{-1}\) as the intensity of transitions involving higher energy states are
too weak to be important. The \(J=0\) matrix blocks produced by TROVE were on
average dimensions \(1\,920\) $\times$ \(1\,920\). The rule of thumb for the
average size of a block for \(J\geq 1\) is \( 1\,920(2J + 1)\). The largest
\(J\) computed was \(J=70\) which required the diagonalisation of matrices in
the order of \(\approx 300\,000\) for eigenvalues and eigenvectors. The linear
algebra libraries LAPACK \citep{99AnBaBi.method} and SCALAPACK \citep{slug}
were employed to solve for the eigenvalues and eigenvectors.

\subsection{Dipole moment surface  and intensities } %Updated DMS

Intensity computation requires high quality electric DMS. We use
an \ai\ DMS computed %by \citet{14YaAlYu.H2CO}
at the CCSD(T)/aug-cc-pVQZ level of theory in the frozen-core approximation
using CFOUR \citep{CFOUR}. Three symmetry-adapted projections of the dipole moment Cartesian
components, $\mu_{\rm A_1}$, $\mu_{\rm B_1}$, and $\mu_{\rm B_2}$, are given in
the analytical representations with each component expanded in Taylor series
(185 parameters in total) in terms of internal coordinates around the
equilibrium configuration using the form developed by \citet{13YaPoTh.H2CS} to
represent the dipole moment of H\2CS.  These parameters reproduce the \ai\
dipole moment values of the $\mu_{\rm A_1}$, $\mu_{\rm B_1}$, and $\mu_{\rm
B_2}$ components with rms errors of 0.0002 Debyes for each component. The
equilibrium value of our dipole moment is 2.3778~D (at $r_{\rm CO}^{\rm e}$ =
1.2033742~\AA, $r_{\rm CH}^{\rm e}$ = 1.10377~\AA, $\theta_{\rm OCH}^{\rm e}$ =
121.844$^{\circ}$),  which can be compared to the experimental value of the
ground vibrational state dipole moment of $\mu$=2.3321(5)~D measured by
\cite{77FaKrMu.H2CO}.

The eigenvectors, obtained by diagonalization, are  used in
conjunction with the DMS to compute the required linestrengths (and
from that the Einstein-$A$ coefficients and absolute intensities) of
transitions that satisfy the rotational selection rules
\begin{equation}\label{e:rot-rules}
   J^{\prime} - J^{\prime\prime} = 0,\pm 1, J^{\prime} + J^{\prime\prime} \ne 0
\end{equation}
and the symmetry selection rules
\begin{equation}\label{e:sym-rules}
 A_{1} \leftrightarrow A_{2}  \;\; , B_{1} \leftrightarrow B_{2}.
\end{equation}
The Einstein-$A$ coefficient for a particular transition from the
\textit{initial} state $i$ to the \textit{final} state $f$ is given by:
\begin{equation}
A_{if} = \frac{8\pi^{4}\tilde{\nu}_{if}^{3}}{3h}(2J_{i} + 1)\sum_{A=X,Y,Z} |\langle \Psi^{f} | \bar{\mu}_{A} | \Psi^{i} \rangle |^{2},
\end{equation}
%\begin{equation}
% S(f \leftarrow i) = g_{\rm ns} \sum_{m_{f}m_{i}} \sum_{A=X,Y,Z} |\langle \Psi^{f} | \bar{\mu_{A}} | \Psi^{i} \rangle |^{2},
%\end{equation}
where $J_{i}$ is the rotational quantum number for the initial state, $h$ is
Planck's constant, \(\tilde{\nu}_{if}\) is the transition frequency (\(hc \,
\tilde{\nu}_{if} = E_{f} -E_{i}\)), \(\Psi^{f}\) and \(\Psi^{i}\) represent the
eigenfunctions of the final and initial states respectively, \(\bar{\mu}_{A}\) is the
electronically averaged component of the dipole moment along the space-fixed
axis \(A=X,Y,Z\) (see also \citet{05YuThCa.method}). From this the absolute
absorption intensity is determined by:
\begin{equation}
I(f \leftarrow i) = \frac{A_{if}}{8\pi c}g_{\rm ns}(2 J_f+1)\frac{\exp\left(-\frac{E_{i}}{kT}\right)}{Q\; \tilde{\nu}_{if}^{2}}\left[1-\exp\left(\frac{-hc\tilde{\nu}_{if}}{kT}\right)\right] ,
% I(f \leftarrow i) = \frac{ 8 \pi^{3} N_{A} \tilde{\nu}_{if}}{(4 \pi \varepsilon_{0}) 3hc} \frac{\exp\left(-\frac{E_{i}}{kT}\right)}{Q}\left[1-\exp\left(\frac{-hc %\tilde{\nu}_{if}}{kT}\right)\right] A_{fi}%(f \leftarrow i),
\label{eq:intens}
\end{equation}
where \(k\) is the Boltzmann constant, \(T\) the absolute temperature and \(g_{\rm ns}\)
is the nuclear spin statistical weight factor. \(Q\), the partition function, is given
by:
\begin{equation}
  Q = \sum_{i} g_{i}\exp\left({\frac{-E_{i}}{k T}}\right),
  \label{eq:part}
\end{equation}
where \(g_{i}\) is the degeneracy of a particular state \(i\) with energy \(E_{i}\). For
H$_{2}$CO, \(g_{i}\) is \(g_{\rm ns}(2 J_i + 1)\) with $g_{\rm ns} = 1$ for \(A_{1}\) and
\(A_{2}\) symmetries and $g_{\rm ns} = 3$  for \(B_{1}\) and \(B_{2}\) symmetries.  The
transitions were computed using the energy limits $hc$ 8\,000 and $hc$ 18\,000~\cm\ for
the lower and upper states, respectively.

Although diagonalisation of the Hamiltonian matrices is very demanding on
computer resources, it is the calculation of the Einstein-$A$ coefficients
which dominates the actual computer time due the sheer number of these and the
large size of the eigenvectors. Graphics processing units (GPU) were therefore
employed to accelerate computation of the intensities. To do this required the
development of a new algorithm to allow these fast but memory poor processors
to be used efficiently. A paper discussing this will be published elsewhere
\citep{jtGAIN}.

\section{Results}

The line list produced, which we call AYTY, contains around 10 billion
transitions with wavenumbers up to 10~000 cm\(^{-1}\). The transitions are
sorted in increasing transition frequency and then converted into the ExoMol
format \citep{jt548}. An extract of the state file and transition file can be
seen in Tables \ref{tab:levels} and \ref{tab:trans}.  Spectra at arbitrary
temperatures can be computed using the Einstein-$A$ coefficients from the
transition files. The theoretical error as estimated by the fitting rms
deviation of 0.18 cm$^{-1}$. This means our transition frequencies and energy
levels should be reliable to about 0.2 cm${-1}$ with low-lying levels,
particularly the pure rotational ones, being much more accurate than this  and
levels for vibrational states for which there are no available laboratory data
much less so.

\begin{table}
\caption{Extract from the H$_2$CO state file. The full table is available at \protect\url{http://cdsarc.u-strasbg.fr/cgi-bin/VizieR?-source=J/MNRAS/}.
} \label{tab:levels}
\begin{center}
\footnotesize
\tabcolsep=5pt
\begin{tabular}{lrcccccccccccccccccccc}
\hline
        $I$  &  \multicolumn{1}{c}{$\tilde{E}$ ,\cm}   &  $g$  &  $J$ & $\Gamma_{\rm tot}$ & $v_1$ & $v_2$ & $v_3$ & $v_4$ & $v_5$ & $v_6$ & $\Gamma_{\rm vib}$ & $K$  & $\Gamma_{\rm rot}$ & $ I_{J,\Gamma}$ & $|C_i^{2}|$ & $n_1$ & $n_2$ & $n_3$ & $n_4$ & $n_5$ & $n_6$ \\
\hline
           1&     0.000000  &    1  &    0  &   1 &  0 &  0 &  0 &  0 &  0 &  0   &   1  &  0  &    1  &      1 &  0.99 &  0 &  0 &  0 &  0 &  0 &  0 \\
           2&  1500.120955  &    1  &    0  &   1 &  0 &  0 &  1 &  0 &  0 &  0   &   1  &  0  &    1  &      2 &  0.92 &  0 &  0 &  0 &  0 &  1 &  0 \\
           3&  1746.045388  &    1  &    0  &   1 &  0 &  1 &  0 &  0 &  0 &  0   &   1  &  0  &    1  &      3 &  0.92 &  1 &  0 &  0 &  0 &  0 &  0 \\
           4&  2327.497142  &    1  &    0  &   1 &  0 &  0 &  0 &  2 &  0 &  0   &   1  &  0  &    1  &      4 &  0.97 &  0 &  0 &  0 &  0 &  0 &  2 \\
           5&  2494.322937  &    1  &    0  &   1 &  0 &  0 &  0 &  0 &  0 &  2   &   1  &  0  &    1  &      5 &  0.96 &  0 &  0 &  0 &  1 &  1 &  0 \\
           6&  2782.410921  &    1  &    0  &   1 &  1 &  0 &  0 &  0 &  0 &  0   &   1  &  0  &    1  &      6 &  0.97 &  0 &  0 &  1 &  0 &  0 &  0 \\
           7&  2999.006647  &    1  &    0  &   1 &  0 &  0 &  2 &  0 &  0 &  0   &   1  &  0  &    1  &      7 &  0.84 &  0 &  0 &  0 &  1 &  1 &  0 \\
           8&  3238.937891  &    1  &    0  &   1 &  0 &  1 &  1 &  0 &  0 &  0   &   1  &  0  &    1  &      8 &  0.70 &  1 &  0 &  0 &  0 &  1 &  0 \\
           9&  3471.719306  &    1  &    0  &   1 &  0 &  2 &  0 &  0 &  0 &  0   &   1  &  0  &    1  &      9 &  0.83 &  2 &  0 &  0 &  0 &  0 &  0 \\
          10&  3825.967015  &    1  &    0  &   1 &  0 &  0 &  1 &  2 &  0 &  0   &   1  &  0  &    1  &     10 &  0.86 &  0 &  0 &  0 &  0 &  1 &  2 \\
          11&  3936.435541  &    1  &    0  &   1 &  0 &  0 &  1 &  0 &  0 &  2   &   1  &  0  &    1  &     11 &  0.73 &  0 &  0 &  0 &  3 &  0 &  0 \\
          12&  4058.101422  &    1  &    0  &   1 &  0 &  1 &  0 &  2 &  0 &  0   &   1  &  0  &    1  &     12 &  0.87 &  1 &  0 &  0 &  0 &  0 &  2 \\
          13&  4083.490190  &    1  &    0  &   1 &  0 &  0 &  0 &  0 &  1 &  1   &   1  &  0  &    1  &     13 &  0.69 &  0 &  1 &  0 &  1 &  0 &  0 \\
          14&  4247.609826  &    1  &    0  &   1 &  0 &  1 &  0 &  0 &  0 &  2   &   1  &  0  &    1  &     14 &  0.79 &  1 &  0 &  0 &  1 &  1 &  0 \\
          15&  4256.314862  &    1  &    0  &   1 &  1 &  0 &  1 &  0 &  0 &  0   &   1  &  0  &    1  &     15 &  0.90 &  0 &  0 &  1 &  0 &  1 &  0 \\
          16&  4495.499848  &    1  &    0  &   1 &  0 &  0 &  3 &  0 &  0 &  0   &   1  &  0  &    1  &     16 &  0.76 &  0 &  0 &  0 &  1 &  2 &  0 \\
          17&  4529.635737  &    1  &    0  &   1 &  1 &  1 &  0 &  0 &  0 &  0   &   1  &  0  &    1  &     17 &  0.90 &  1 &  0 &  1 &  0 &  0 &  0 \\

\hline
\end{tabular}
\end{center}
\noindent
$I$:   State counting number; \\
$\tilde{E}$: State term energy in \cm; \\
$g$: State degeneracy; \\
$J$:   State rotational quantum number; \\
$\Gamma_{\rm tot}$: Total symmetry in $C_{2\nu}(M)$ (1 is $A_1$, 2 is $A_2$, 3 is $B_1$ and 4 is $B_2$ ); \\
$v_1 - v_6$:  Normal mode vibrational quantum numbers; \\
$\Gamma_{\rm vib}$: Symmetry of vibrational contribution in $C_{2\nu}(M)$; \\
$K$:   State projection of the rotational quantum number; \\
$\Gamma_{\rm rot}$: Symmetry of rotational contribution in $C_{2\nu}(M)$ ;  \\
$I_{J,\Gamma}$: State number in $J,\Gamma$ block; \\
$|C_i^{2}|$:  Largest coefficient used in the assignment;\\
$n_1 - n_6$:  TROVE vibrational quantum numbers.\\
\end{table}

\begin{table}
\caption{Extracts from the H$_2$CO transitions file.
The full table is available at \protect\url{http://cdsarc.u-strasbg.fr/cgi-bin/VizieR?-source=J/MNRAS/}}
\label{tab:trans}
\begin{center}
\begin{tabular}{rrr}
\hline
       \multicolumn{1}{c}{$f$}  &  \multicolumn{1}{c}{$i$} & \multicolumn{1}{c}{$A_{fi}$}\\
\hline

     6713828    &     6734990  &  8.2910e-06 \\
     6709468    &     6722660  &  3.2621e-05 \\
     6704996    &     6726710  &  4.7333e-05 \\
     6726711    &     6739070  &  5.0697e-05 \\
     6718218    &     6730865  &  5.4273e-05 \\
     6730866    &     6750469  &  5.6752e-05 \\
\hline
\end{tabular}

\noindent
 $f$: Upper  state counting number;\\
$i$:  Lower state counting number; \\
$A_{fi}$:  Einstein-$A$ coefficient in s$^{-1}$.

\end{center}
\end{table}

The completeness of the line list as a function of temperature can be
determined by checking the convergence of the temperature-dependent partition
function \(Q\) given in Eq.~(\ref{eq:part}), which is computed via explicit
summation \citep{jt263} of the 10.3 million energy levels available. As \(T\)
increases, a greater proportion of these states are required as their
contribution towards \(Q\) becomes more important. Figure \ref{fig:part_plot}
shows our computed partition function as a function of the maximum $J$ value
(\(J_{\rm max}\)) used in the calculation. As \(J_{\rm max}\) increases, each
$J$ contributes progressively less until convergence is reached. The partition
function at $T=296$ K converges to better than 1\% at \(J \approx 34\) with the
limit of \(Q=2\,844.621\) at \(J=58\).  For $=1\,500$ K, it converges to about
0.005\% at \(J=70\) with a \(Q\) value of 130~190.25. These partition functions
can be used to evaluate the effect of lower energy state threshold of 8000
cm\(^{-1}\) on the completeness of the line list by comparing $Q_{\rm
  limit}$, which sums energies up to this threshold, with the full
partition sum.  Figure \ref{fig:part_plot_limit} shows that the two
partition functions are essentially the same up to 800~K and that
$Q_{\rm limit}$ is 92.3 \% of $Q$ at $T= 1500$~K.  Therefore we
recommend $T$=1500~K as a `soft' limit to the applicability of the
line list.  Use of the line list at higher temperatures will lead to
the progressive loss of opacity although the ratio $Q_{\rm limit}/Q$ can
be used to estimate the proportion of this missing contribution
\citep{jt181}.

%\red{CAN WE COMPARE WITH HIGH T Q OF H2CO BY Martin ET AL J. Chem. Phys. 95, 8374 (1991)?}
%\blue{Martin ET AL IS ROTATIONAL ONLY. I TRIED TO COMPUTE THEM FOR ALL OF THE TEMPERATURES IN THE PAPER USING THE PURE ROTATIONAL STATES BUT I STILL DONT GET THE RIGHT NUMBERS.}

Table \ref{tab:part} compares our partition functions with those from CDMS
\citep{05MuScSt.db} and those used in HITRAN \citep{03FiGaGo}.  At temperatures $T \leq
300$~K we agree to better than \(1\%\) with CDMS and HITRAN. At 500~K the
difference with CDMS is much higher at 8.9\%, due our explicit sum running over
a much larger number levels, but agreement with HITRAN is good. There are
bigger differences at higher temperatures: at 1500~K our partition
function is lower by about 1.2\%\ and at 3000~K by 9.7\%. This may be caused by
the lack of the high energy contributions due to the energy cut-off of $hc \,
18\,000$~\icm\ used in our line list, see  \citet{jt571} and \citet{jt169} for a discussion of the importance
of contributions from the excited ro-vibrational states up to the dissociation.
Our full partition function evaluated on a 1 K grid is given in the
supplementary data.

We use the analytical representation suggested by \citet{jt263} as given by
\begin{equation}
\log_{10} Q(T) = \sum_{n=0}^8 a_n \left[\log_{10} T\right]^n \label{eq:pffit}.
\end{equation}
The expansion parameters  given in Table~\ref{tab:pffit} reproduce our
partition function better than 0.3\%\ for temperatures ranging up to 3000~K.

%Maybe add why ours is good
\iffalse
\begin{figure}[!t]
  \begin{subfigure}
    \centering
    \epsfxsize=8.0cm \epsfbox{part_296K.eps}
    %\includegraphics[width=100mm]{part_296K.eps}
    \caption{$T$=296K}
    \label{fig:part_296K}
  \end{subfigure}
  \begin{subfigure}
    \centering
    \epsfxsize=8.0cm \epsfbox{part_1500K.eps}
    %\includegraphics[width=100mm]{part_1500K.eps}
    \caption{$T$=1500K}
    \label{fig:part_1500K}
  \end{subfigure}
  \label{fig:part_plot}
  \caption{Partition function plots against inclusion of \(J_{\rm max}\) states.}
\end{figure}
\fi
\begin{figure}[!t]
\centering
\epsfxsize=14.0cm \epsfbox{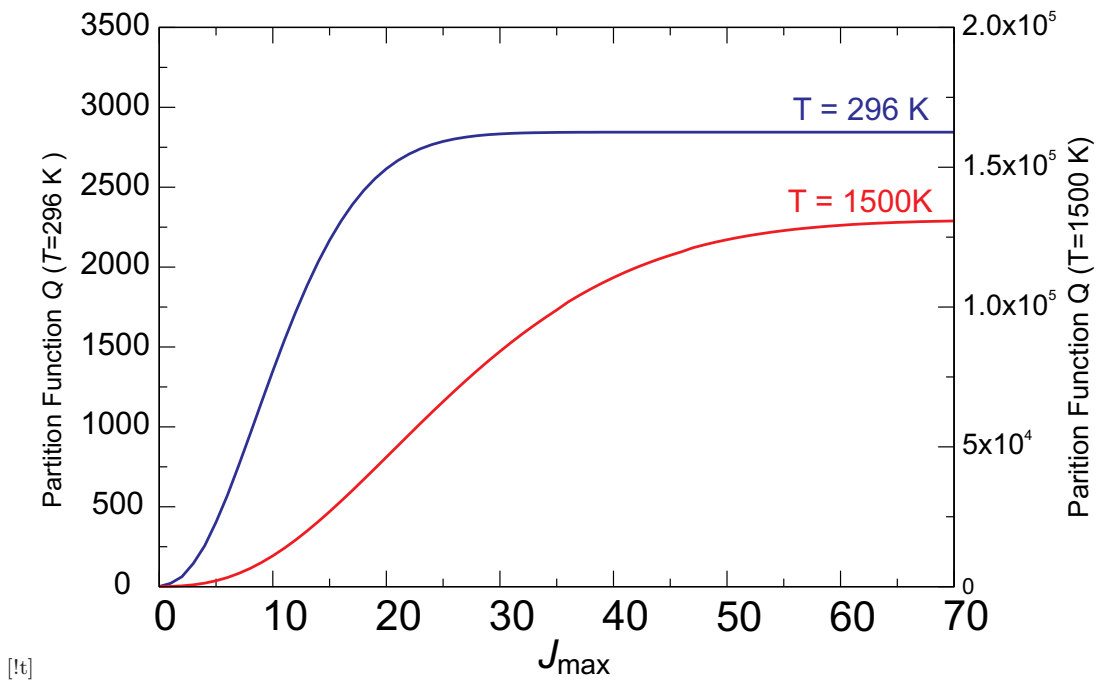}
\caption{Partition functions at two temperatures as a function of inclusion of rotational
states: all $J$ up to \(J_{\rm max}\) for $T=296$ K (left hand scale) and $T=1500$ K (right hand scale).}
\label{fig:part_plot}
\end{figure}
\begin{table}
\caption{Parameters used to represent the partition function,
see Eq.~(\ref{eq:pffit}), valid for temperatures up to 3000 K.}
\label{tab:pffit} \footnotesize
\begin{center}
\begin{tabular}{lr}
\hline
Parameter &  Value  \\
\hline
$	a_0	$&$	1.12789807683	$\\
$	a_1	$&$	-5.35067939866	$\\
$	a_2	$&$	10.33684323700	$\\
$	a_3	$&$	-4.92187455147	$\\
$	a_4	$&$	-2.28234089365	$\\
$	a_5	$&$	3.61122821799	$\\
$	a_6	$&$	-1.64174365325	$\\
$	a_7	$&$	0.33727543206	$\\
$	a_8	$&$	-0.02654223136	$\\
\hline
\end{tabular}
\end{center}
\end{table}

\begin{figure}
\centering
\epsfxsize=14.0cm \epsfbox{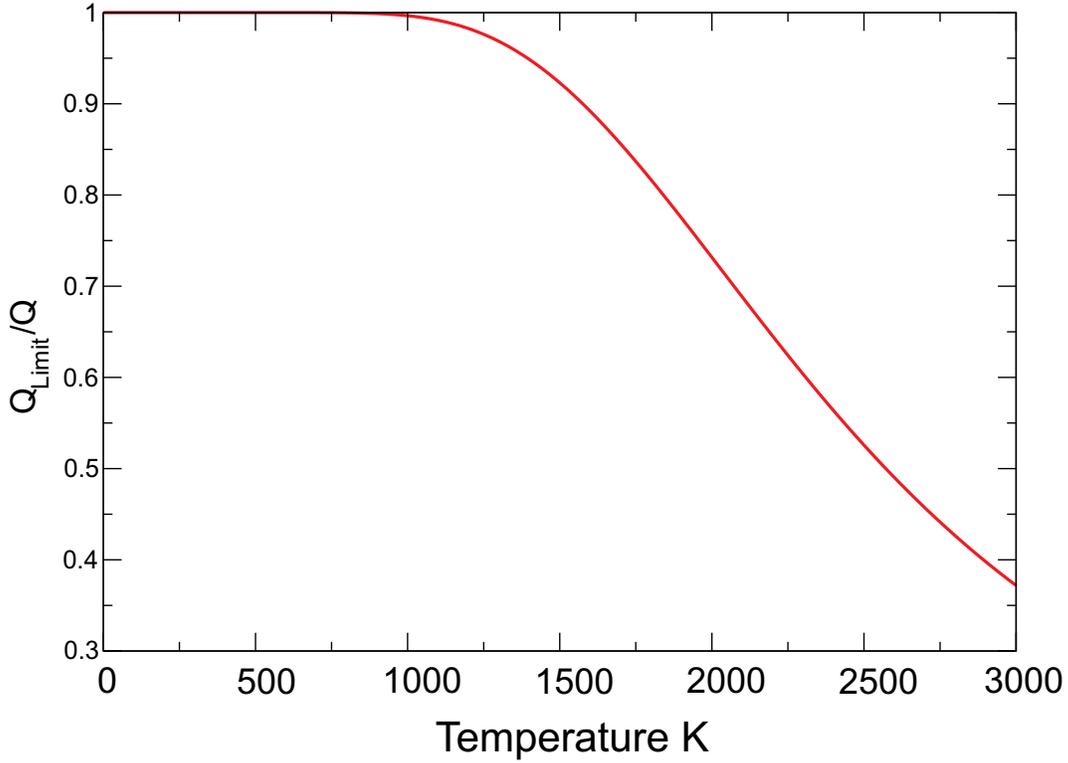}
\caption{Plot of \(Q_{\rm limit}\)/\(Q\) against temperature where \(Q_{\rm limit}\) is the partition function
computed using only energy levels below our lower state threshold of 8000 cm\(^{-1}\).}
\label{fig:part_plot_limit}
\end{figure}

\begin{table}
\caption{Comparisons of H\2CO partition functions as a function of
temperature for this work, CDMS \citep{05MuScSt.db} and those used in HITRAN \citep{03FiGaGo}.}
\begin{center}
\begin{tabular}{crrrr}
\hline\hline
$T$ / K         & AYTY     & CDMS      & HITRAN \\
\hline
2.725 & 2.0165    & 2.0166    \\
5.000 & 4.4833    & 4.4832    & \\
9.375 & 13.801    & 13.8008   & \\
18.75 & 44.6835   & 44.6812   & \\
37.5 & 128.6581  & 128.6492  & \\
75 & 361.7053  & 361.7195  &  362.07\\
150 & 1019.9549 & 1019.9706 &  1020.47\\
225& 1874.4679 & 1872.6221 &  1875.67\\
300& 2904.1778 & 2883.0163 &  2906.32\\
500& 6760.2315 & 6208.3442 & 6760.99\\
1500 & 128635.40&& 130190.25 \\
3000 & 2741283.3 && 3038800.0 \\
\hline
\end{tabular}
\end{center}
\label{tab:part}
\end{table}
\begin{figure}
\centering
\epsfxsize=14.0cm \epsfbox{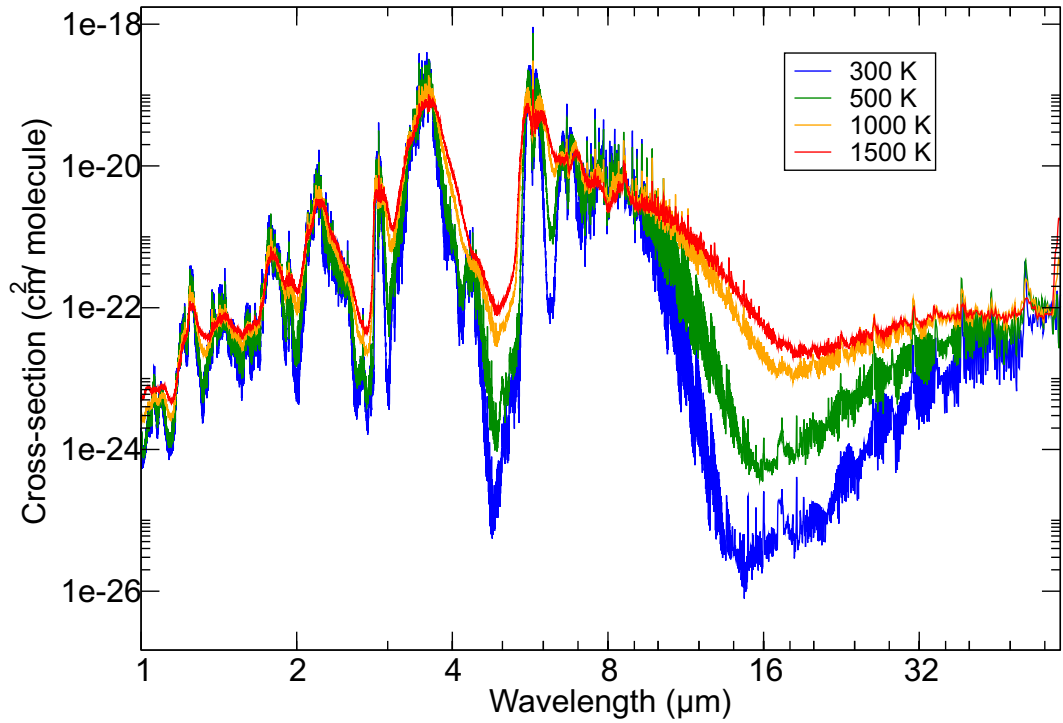}
\caption{ Cross-sections of the entire AYTY line list as a function of temperature: The curves
in the 16~$\mu$m region increase in opacity with increasing temperature.}
\label{fig:trove_dep}
\end{figure}
\begin{figure}
    \centering
\epsfxsize=14.0cm \epsfbox{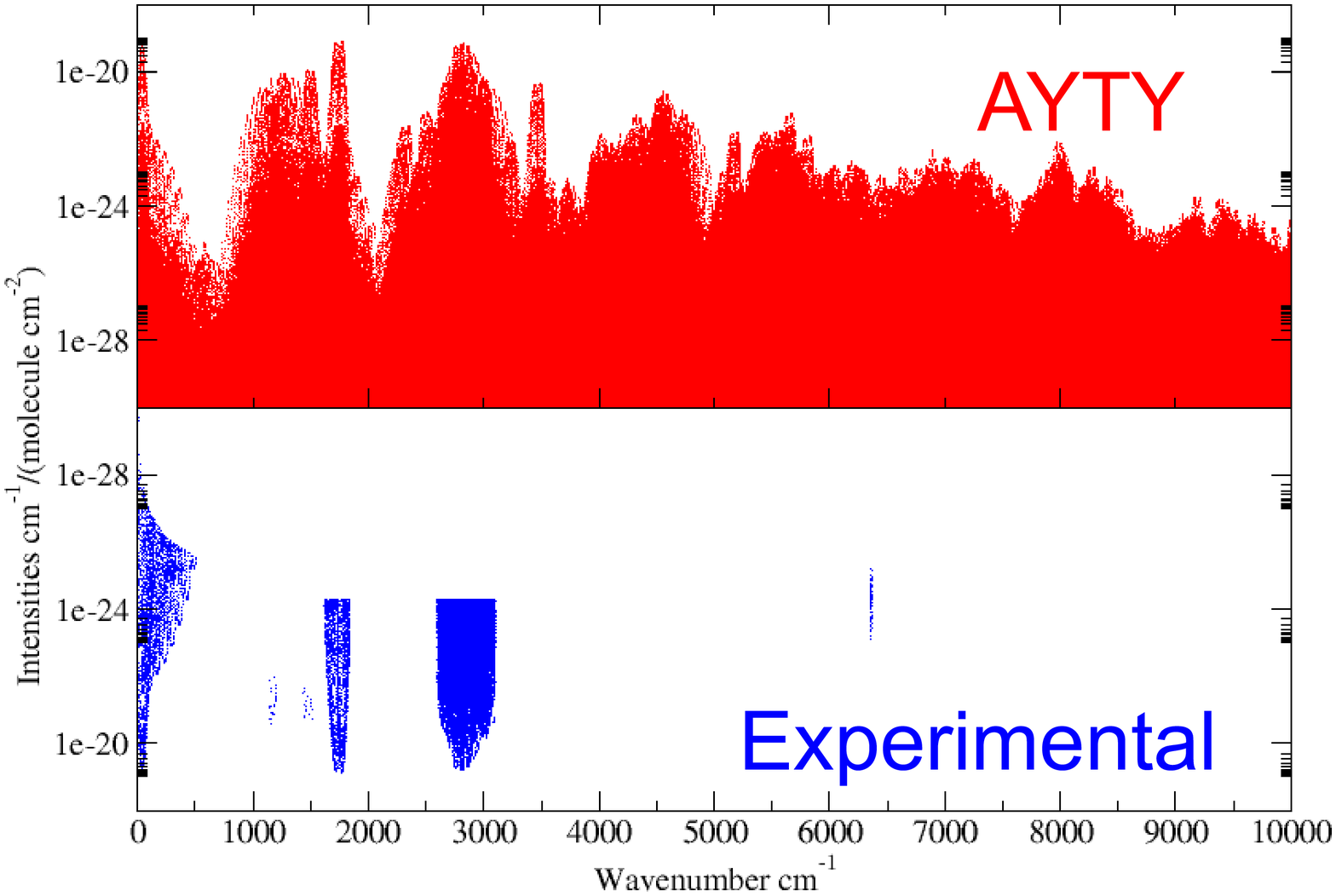}
\caption{Overview of our synthetic spectrum at $T=$296 K against HITRAN \protect\citep{jt557}, \citet{89ReNaDa.H2CO} and \citet{07ZhGaDe.H2CO}.}
%(0-550 cm$^{-1}$, 1600-3100 cm$^{-1}$), \protect\cite{03PeKeFl.H2CO} (1148-1517 cm$^{-1}$) and \protect\cite{07ZhGaDe.H2CO} (6351-6363 cm$^{-1}$).
    \label{fig:trove_hi_log}
\end{figure}

\begin{figure}
\centering
%\red{png for now, trying to reduce size of eps}
\epsfxsize=14.0cm \epsfbox{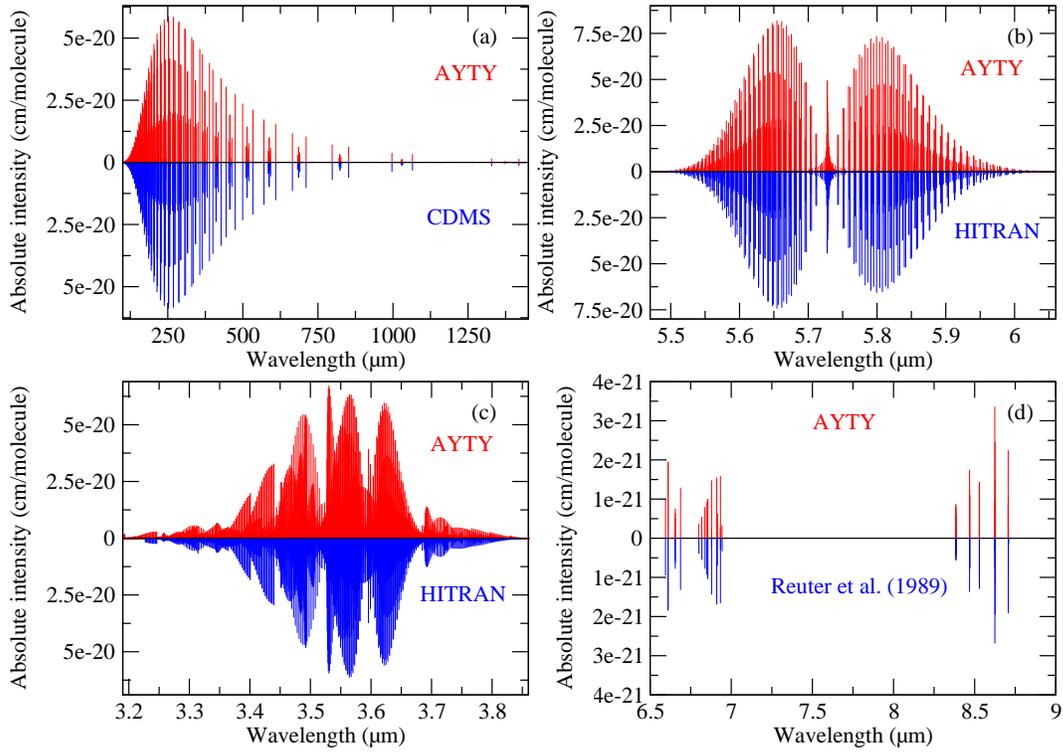}
\caption{The fundamental bands compared to currently available experimental intensities
\citep{jt557,89ReNaDa.H2CO,05MuScSt.db} at $T=296$ K. (a) Rotational Band (b) \(\nu_{2}\) (c) \(\nu_{1}\)
and \(\nu_{5}\) (d) \(\nu_{3}\), \(\nu_{4}\) and \(\nu_{6}\) . }
\label{fig:trove_3figs}
\end{figure}

The dependance of the cross-sections on temperature is illustrated in Figure
\ref{fig:trove_dep}, the features in the simulated spectra become smoother as
the temperature increases. This is a result of the vibrationally excited states
becoming more populated and the increasing width of the rotational envelope
with temperature. Figure \ref{fig:trove_hi_log} shows a simulated $T=296$~K
spectrum computed from our line list against the available laboratory
absorption spectra up to 10~000 cm\(^{-1}\).The logarithmic scale used shows
the density of transitions in our line list and reveals the significant gaps
and limitations in the HITRAN 2012 database. Comparing specific regions, our
line list accurately replicates both the line positions and intensities of the
three available bands, as illustrated in detail in Fig.~\ref{fig:trove_3figs}.
Additional lines are present as our computed spectra contains all possible
transitions within the region including hot bands. Fig.~\ref{fig:trove_3figs}d
and Table \ref{tab:missingbandv6v4v3} show agreement with the line positions
and absoulte intensities from \citet{89ReNaDa.H2CO} with an rms deviation of
0.099 cm$^{-1}$ for the line positions. There are some limitations with our
line list. Higher \(J\) transitions at around the \(J> 50\) range begin to show
a slight drift of \(\approx 0.3\) cm\(^{-1}\) in predicted line position; this
does not occur for the rotational band. In practice, errors in the
ro-vibrational energy levels grow with $K$ (as opposed to $J$); the
discrepancies in transition frequencies become more pronounced in $|K'-K''|=1$
transitions than those that involve the same $K$ ($K\p=K\pp$). This can be seen
in the lack of drift in the pure rotational band as it is mostly comprised of
$K\p = K\pp$ transitions due to both ground and excited states being of $A_1$
symmetry. $B_1$ and $B_2$ vibrational bands however are mostly comprised of
$|K\p-K\pp|=1$ transitions which makes their errors more sensitive to the
quality of the model.
%to the truncations used in our computation or

%\red{CAN WE COMPARE WITH \url{http://dx.doi.org/10.1063/1.471537?} and  with
%\url{http://dx.doi.org/10.1080/00268970600641568?} and with
%\url{http://pubs.rsc.org/en/Content/ArticleLanding/2002/CP/b109300m#!divAbstract} (Phys.
%Chem. Chem. Phys., 2002,4, 445-450)}

\begin{figure}
\centering
%\red{png for now, trying to reduce size of eps}
\epsfxsize=14.0cm \epsfbox{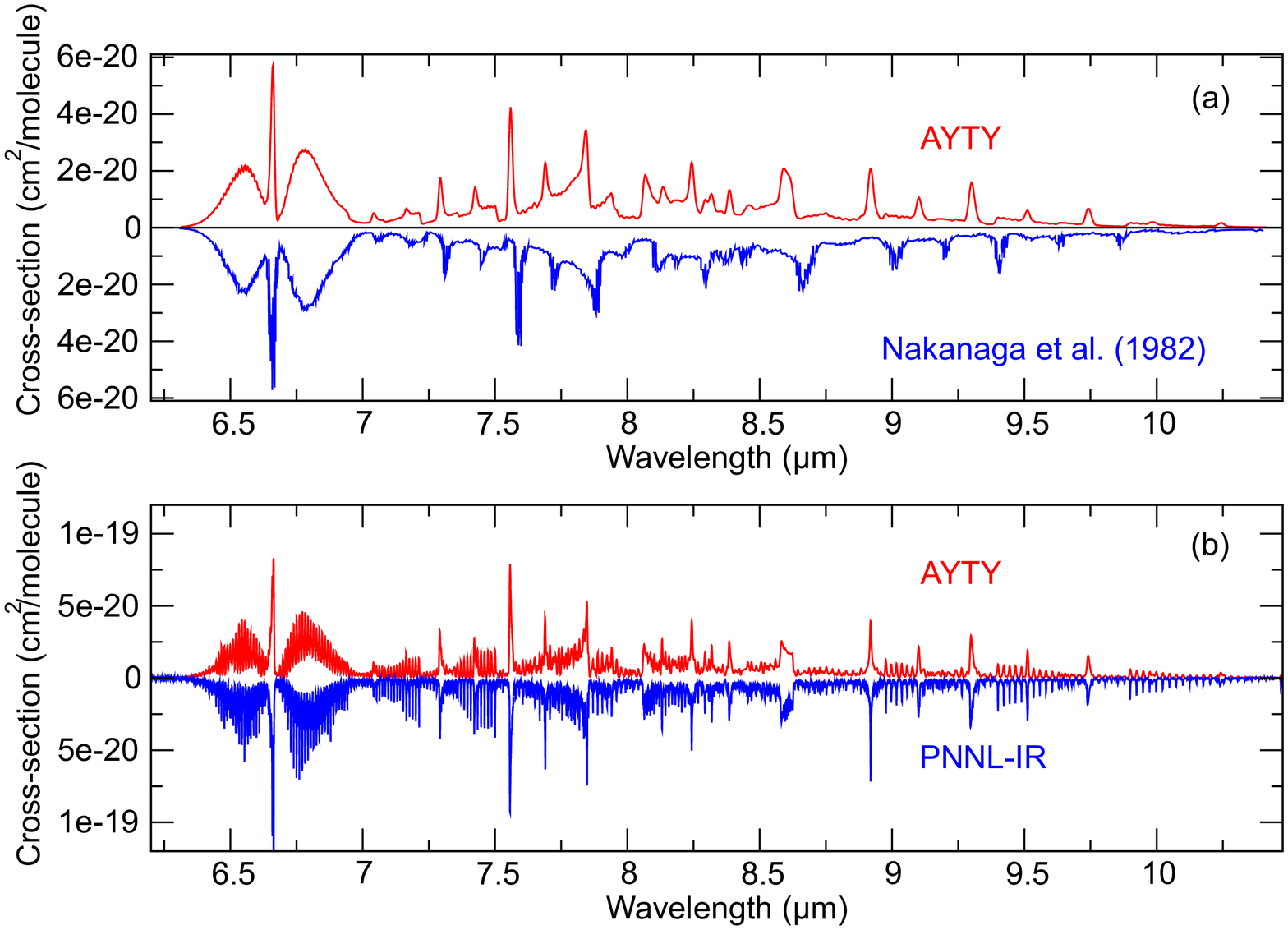}
\caption{Cross-section comparison of AYTY against experimental of the \(\nu_{3}\), \(\nu_{4}\) and \(\nu_{6}\) band regions:
(a) \citet{82NaTaKo.H2CO} at 296 K with HWHM = 1.1849 cm$^{-1}$;
Extracted from image and scaled to match the AYTY cross-section; (b) PNNL-IR data at 323.15 K \citep{PNNL} with
HWHM = 0.1120 cm$^{-1}$. }
\label{fig:v3v4v6cross}
\end{figure}

Computing band intensities requires simulating spectra at a chosen temperature
and accumulating all transitions that correspond to the specific band.
Table~\ref{tab:bandintens} highlights our band intensities against those
available in the literature. Each band intensity required spectra simulated to
the parameters used by each reference. In general, AYTY agrees well with all
band intensities but is more intense. This may be due to the fact that AYTY
sums over all lines in a given band whereas experiments generally only capture
the strongest lines. Table~\ref{tab:bandintens} also shows the total band
intensity for the 3.5~$\mu$m region compared to that by
\cite{79BrHuPi.H2CO,82NaTaKo.H2CO} and HITRAN. Our value is 13~\%\ stronger
than HITRAN, (matches the discrepancy for the  $\nu_1$ and $\nu_5$ bands  in
Table~\ref{tab:bandintens}), 18~\%\ stronger than \citet{82NaTaKo.H2CO} and
40~\%\ stronger than that by \citet{79BrHuPi.H2CO}. Absolute intensities and
bands not available in the HITRAN database or literature can be evaluated
against cross-sections. For the $\nu_3$, $\nu_4$ and $\nu_6$ bands, further
evaluation of these bands can be made against cross-sections available from the
PNNL-IR database \citep{PNNL} and \citet{82NaTaKo.H2CO} using a Gaussian
profile with a HWHM (half-width-half-maximum) of 1.1849 cm$^{-1}$ and 0.1120
cm$^{-1}$, determined from their respective experimental profiles. Figure
\ref{fig:v3v4v6cross}a compares the AYTY line list with a spectrum extracted
from Fig.~3 of \citet{82NaTaKo.H2CO} and scaled to match the AYTY line list.
Good agreement is seen in both structure and position in the band with a slight
drift occurring as an artifact from the extraction process. Figure
\ref{fig:v3v4v6cross}b shows an even better agreement with the spectral
structure as well as the cross-section intensity.

\begin{figure}
\centering
\epsfxsize=14.0cm \epsfbox{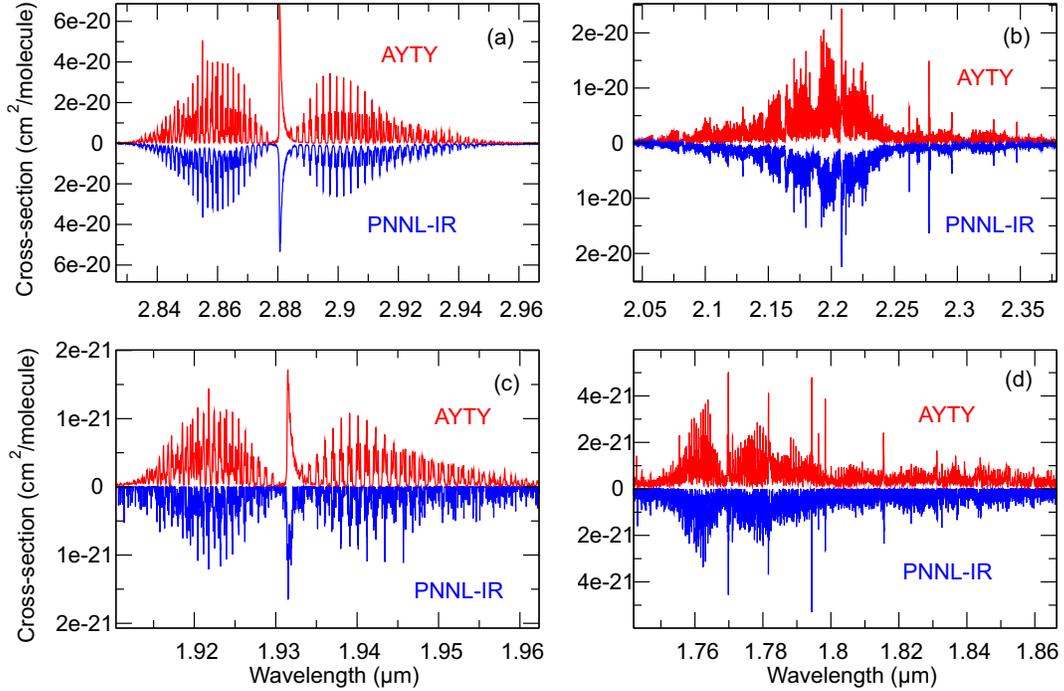}
\caption{Additional bands in PNNL at wavelengths below 3.2 $\mu m$ with HWHM at 0.1120 \icm. (a) \(2\nu_{2}\) band;
(b) Bands covered by \citet{06FlLaSa.H2CO}; (c) \(3\nu_{2}\);
(d) Various bands including \(\nu_{1}+\nu_{5}\).
 Note: (c) and (d) Negative PNNL values have been removed.}
\label{fig:morebands}
\end{figure}

The total integrated cross-section over the region 6.2
 -- 10.5 $\mu$m for AYTY and PNNL is \(8.02\times10^{-17}\)
cm/molecule and \(8.20\times10^{-17}\) cm/molecule respectively, making PNNL
overall around 8\% stronger. PNNL covers regions beyond those currently
available in HITRAN. Figure \ref{fig:morebands}(a) depicts the $2\nu_{2}$ band
at 2.88 \micro. Good agreement is seen in structure, position and
cross-sections with the integrated cross-sections differing by only 10\%.
%This correlates
%with the measured $2\nu_{2}$ transition moment of 2.3405 Debye from
%\citet{75JoMcxx.H2CO} and our computed transition moment of 2.352 agreeing to better than 1\%.
The regions below 2.8 \micro~in PNNL become increasingly
polluted with noise but band features are still visible as seen in Figures
\ref{fig:morebands}(b),(c) and (d). In particular, Figure
\ref{fig:morebands}(b), the AYTY cross-section reproduces peaks in features
present in the PNNL-IR data. This region was also studied by
\citet{06FlLaSa.H2CO}. Their absorbance spectrum produces certain transitions
with double the intensity compared to AYTY. These are due to splitting caused by
two transitions with the same quanta but with swapped $\Gamma_{f}$ and
$\Gamma_{i}$ giving the two lines very similar transition frequencies and absolute
intensity which make them difficult to resolve experimentally.

%3.5~$\mu$m band & 776.8 &  682.7 &  546.9                            653.5

Further bands include the integrated cross-section for the \(2\nu_{5}\) band at
5676.21 cm$^{-1}$ for AYTY and \citet{02BaCoHa.H2CO} at \(6.4\times10^{-22}\)
cm/molecule and \(5.6\times10^{-22}\) cm/molecule respectively making AYTY 11\%
stronger. In Table~\ref{tab:bandintens} we also compare theoretical (AYTY)
overtone band intensities obtained by the direct summation with the
corresponding experimentally derived values from
\citet{06PeVaDa.H2CO,06FlLaSa.H2CO}. The agreement with the data obtained by
\citet{06FlLaSa.H2CO} is very good. Those from \citet{06PeVaDa.H2CO} are in
fact a compilation of different sources
\citep{55HiEgxx.H2CO,79BrHuPi.H2CO,82NaTaKo.H2CO,05HeNeLi.H2CO,88ClVaxx.H2CO},
some of which were obtained at low resolution, which could explain the
%less good agreement with our estimations.
slightly worse agreement with our calculations. Compare the total total
integrated band intensity for the band at 1.5~$\mu$m we obtain $3.11\times
10^{-17}$ cm$/$molecule against 2.19, 2.62, and 2.73 $\times 10^{-17}$ cm$/$
molecule by \citet{06PeVaDa.H2CO}, \citet{82NaTaKo.H2CO} and from HITRAN,
respectively.

Finally, \citet{94ItNaTa.H2CO} presented the relative band intensities as the
ratio of the vibrational transition moments between $2\nu_4$ and $2\nu_6$ of
0.755(48), which can be compared to our absolute value of 0.6264.

%http://www.sciencedirect.com/science/article/pii/0584853994E0050K#  computes a
%vibrational tm ratio for 2v4/2v6 of -0.755(48) (or 0.52(12) as another), ours
%is 0.6264.

\begin{table}
 \caption{Band intensities, in 10\(^{-17}\) cm$^{-1}$/(molecule cm$^{-2}$).
%for formaldehyde. \red{CAN WE COMPARE OTHER BAND INTENSITIES AS WELL? FOR EXAMPLE,} \protect\url{http://dx.doi.org/10.1063/1.431497};
%\red{JOURNAL OF MOLECULAR SPECTROSCOPY 67, 476-495 (1975)}}
%\red{I made a comparison with the specific 2v2 state highlighted in the paper when discussing the PNNL-region but do you want a comparison against all of the transition moments? Wouldnt it be more relevant in the DMS part of the paper?}
}
\begin{center}
\begin{threeparttable}
\begin{tabular}{lcllr}
%ALL OF OUR BANDS HAVE IMPROVED SIGNIFICANTLY YAY!!!!
\hline
Band & Ref. & Obs & AYTY & (O-A)/O (\%)      \\
\hline
\(\nu_{1}\)                  & \tnote{a}    & 1.008               & 1.057               & -4.9  \\
\(\nu_{2}\)                  & \tnote{a}    & 1.219               & 1.348                & -10.6  \\
\(\nu_{3}\)                  & \tnote{b}    & 0.184             & 0.185               & -0.5  \\
\(\nu_{4}\)                  & \tnote{b}    & 0.069             & 0.089               & -27.8 \\
\(\nu_{5}\)                  & \tnote{a}    & 1.120                & 1.282                 & -14.6     \\
\(\nu_{6}\)                  & \tnote{a}    & 0.173              & 0.204                & -17.9    \\
  $  \nu_2+\nu_3           $ & \tnote{c}    &                0.0025  &                0.0019  &        22.7\\
  $  \nu_2+\nu_6           $ & \tnote{c}    &                0.0790  &                0.1222  &       -54.6\\
  $  2\nu_3                $ & \tnote{c}    &                0.0260  &                0.0428  &       -64.5\\
  $  \nu_2+\nu_4           $ & \tnote{c}    &                0.1100  &                0.1379  &       -25.4\\
  $  \nu_3+\nu_6           $ & \tnote{c}    &                0.1940  &                0.3274  &       -68.8\\
  $  \nu_3+\nu_4           $ & \tnote{c}    &                0.0290  &                0.0300  &        -3.4\\
  $  2\nu_6                $ & \tnote{c}    &                0.0220  &                0.0214  &         2.9\\
  $  \nu_4+\nu_6           $ & \tnote{c}    &                0.0062  &                0.0014  &        77.7\\
  $  2\nu_4                $ & \tnote{c}    &                0.0060  &                0.0047  &        22.4\\
  $  \nu_1+\nu_6           $ & \tnote{d}    &                0.0015  &                0.0022  &       -45.0\\
  $  \nu_2+\nu_4+\nu_6     $ & \tnote{d}    &                0.0006  &                0.0007  &        -4.6\\
  $  \nu_3+\nu_5           $ & \tnote{d}    &                0.0097  &                0.0098  &        -1.2\\
  $  2\nu_3+\nu_6          $ & \tnote{d}    &                0.0036  &                0.0027  &        24.4\\
  $  \nu_2+\nu_5           $ & \tnote{d}    &                0.0377  &                0.0446  &       -18.2\\
  $  2\nu_2+\nu_6          $ & \tnote{d}    &                0.0108  &                0.0123  &       -14.0\\
  $  \nu_1+\nu_2           $ & \tnote{d}    &                0.0243  &                0.0275  &       -13.2\\
  $  3\nu_2                $ & \tnote{d}    &                0.0022  &                0.0026  &       -21.4\\
\hline
\end{tabular}
\end{threeparttable}
\begin{center}
 \begin{tablenotes}
 \item [a] \citet{09PeJaTc.H2CO}
 \item [b] \citet{03PeKeFl.H2CO}
  \item [c] \citet{06PeVaDa.H2CO}
 \item [d] \citet{06FlLaSa.H2CO}
\end{tablenotes}
\end{center}
\end{center}
\label{tab:bandintens}
\end{table}

\begin{table}
  \caption{Residuals, in cm\(^{-1}\),  for line positions for the $\nu_3$,
$\nu_4$ and $\nu_6$ bands. Observed data from \citet{89ReNaDa.H2CO}. }
%Data only came from Reuter is it worth including DangNhu?
  \centering
\begin{tabular}{crrllr}
\hline\hline
Band & $J^{\prime}$ & $J^{\prime\prime}$ & AYTY & Obs. & Obs.-Calc. \\
\hline
6 & 17 & 18 & 1148.4322 & 1148.3346 & -0.0976 \\
6 & 17 & 18 & 1148.4578 & 1148.3600 & -0.0978 \\
4 & 11 & 10 & 1148.4115 & 1148.3453 & -0.0662 \\
4 & 3  & 4  & 1148.5548 & 1148.4702 & -0.0846 \\
4 & 16 & 16 & 1148.6150 & 1148.5082 & -0.1068 \\
4 & 1  & 1  & 1159.2222 & 1159.1356 & -0.0866 \\
4 & 2  & 2  & 1159.3587 & 1159.2716 & -0.0871 \\
4 & 28 & 28 & 1159.3222 & 1159.3070 & -0.0152 \\
4 & 15 & 14 & 1159.4760 & 1159.3917 & -0.0843 \\
6 & 6  & 7  & 1159.5539 & 1159.4132 & -0.1407 \\
4 & 9  & 8  & 1159.5115 & 1159.4396 & -0.0719 \\
4 & 3  & 3  & 1159.5594 & 1159.4715 & -0.0879 \\
4 & 18 & 18 & 1172.4242 & 1172.3864 & -0.0378 \\
4 & 6  & 6  & 1172.6002 & 1172.5256 & -0.0746 \\
6 & 12 & 13 & 1180.6607 & 1180.6446 & -0.0161 \\
6 & 4  & 5  & 1180.8080 & 1180.7328 & -0.0752 \\
4 & 24 & 23 & 1180.8209 & 1180.8082 & -0.0127 \\
4 & 11 & 11 & 1180.8777 & 1180.8324 & -0.0453 \\
6 & 13 & 14 & 1180.9109 & 1180.8834 & -0.0275 \\
6 & 10 & 10 & 1192.6923 & 1192.6086 & -0.0837 \\
6 & 3  & 4  & 1192.6678 & 1192.6267 & -0.0411 \\
6 & 9  & 9  & 1192.7477 & 1192.6657 & -0.0820 \\
6 & 8  & 8  & 1192.7985 & 1192.7181 & -0.0804 \\
6 & 18 & 19 & 1192.7781 & 1192.7369 & -0.0412 \\
6 & 7  & 7  & 1192.8441 & 1192.7651 & -0.0790 \\
6 & 10 & 11 & 1192.7723 & 1192.7954 & 0.0231  \\
6 & 6  & 6  & 1192.8845 & 1192.8067 & -0.0778 \\
6 & 5  & 5  & 1192.9194 & 1192.8427 & -0.0767 \\
6 & 19 & 18 & 1440.3351 & 1440.1330 & -0.2021 \\
6 & 20 & 19 & 1442.4701 & 1442.2633 & -0.2068 \\
6 & 17 & 16 & 1460.4035 & 1460.1831 & -0.2204 \\
6 & 18 & 17 & 1462.5117 & 1462.2863 & -0.2254 \\
6 & 20 & 19 & 1466.6771 & 1466.4415 & -0.2356 \\
6 & 22 & 21 & 1470.7763 & 1470.5289 & -0.2474 \\
3 & 24 & 25 & 1442.2597 & 1442.2329 & -0.0268 \\
3 & 22 & 23 & 1446.1622 & 1446.2088 & 0.0466  \\
3 & 22 & 23 & 1447.1125 & 1447.1250 & 0.0125  \\
3 & 19 & 20 & 1453.7010 & 1453.7073 & 0.0063  \\
3 & 20 & 21 & 1453.7236 & 1453.7154 & -0.0082 \\
3 & 17 & 18 & 1458.7043 & 1458.7231 & 0.0188  \\
3 & 17 & 18 & 1458.7141 & 1458.7325 & 0.0184  \\
3 & 1  & 2  & 1495.2715 & 1495.3254 & 0.0539  \\
3 & 1  & 0  & 1502.5583 & 1502.6118 & 0.0535  \\
3 & 9  & 9  & 1502.6600 & 1502.6548 & -0.0052 \\
3 & 11 & 11 & 1502.9248 & 1502.9188 & -0.0060 \\
3 & 5  & 4  & 1512.6148 & 1512.6595 & 0.0447  \\
3 & 5  & 4  & 1512.6741 & 1512.7189 & 0.0448  \\
3 & 6  & 5  & 1516.3213 & 1516.3326 & 0.0113  \\
\hline
\end{tabular}
  \label{tab:missingbandv6v4v3}
\end{table}

\section{Conclusion}

We have computed the frequency and Einstein-$A$ coefficients of almost 10
billion transitions of formaldehyde, which cover wavelengths longer that 1
$\mu$m and includes all rotational excitations up to \(J=70\), making
the line list applicable for temperatures  up to 1500 K. The AYTY line list gives a
room-temperature spectrum in excellent agreement with available experimental
data. We have highlighted those regions missing from the HITRAN database with
the hope that they will be investigated further experimentally. The new line
list may be accessed via \url{www.exomol.com} or
\url{http://cdsarc.u-strasbg.fr/viz-bin/qcat?J/MNRAS/}. The cross-sections of
H\2CO can be also generated at \url{www.exomol.com} as described by
\citet{jt542}.

\section{Acknowledgements}
This work was supported by the ERC under the Advanced Investigator
Project 267219 and made use of the DiRAC@Darwin, DiRAC@COSMOS HPC cluster and Emerald CfI cluster. DiRAC is the UK HPC facility for
particle physics, astrophysics and cosmology and is supported by STFC and BIS. The authors would like to acknowledge the work presented here made use of the EMERALD High Performance Computing facility provided via the Centre for Innovation (CfI). The CfI is formed from the universities of Bristol, Oxford, Southampton and UCL in partnership with STFC Rutherford Appleton Laboratory. We thank Clara Sousa-Silva and Duncan A.
Little for help during the writing of this paper, AFA would
also like to thank Dr. Faris N. Al-Refaie, Lamya Ali, Sarfraz Ahmed Aziz, and Rory and Annie Gleeson for their support.

% Uncomment the following two lines if you want to have a bibliography
\bibliographystyle{mn2e}
%\bibliography{journals_astro,H2CO,linelists,methods,exogen,additional,jtj,H2CS,NH3,SbH3,HSOH,abinitio}

\label{lastpage}

\end{document}